\newcommand{\be}{\begin{equation}}      
\newcommand{\ee}{\end{equation}}
\newcommand{\bea}{\begin{eqnarray}}     
\newcommand{\eea}{\end{eqnarray}}
\newcommand{\beb}{\begin{eqnarray*}}    
\newcommand{\eeb}{\end{eqnarray*}}

\documentclass[twocolumn,,floatfix,showpacs,prl]{revtex4}
 \usepackage{graphicx}

\begin{document}
\title{Polarization transitions in quantum ring arrays}

\author{Bahman Roostaei$^{1}$}
\altaffiliation[Also at ]{Center for Semiconductor Physics in Nanostructures}

\author{K. Mullen$^1$}
\altaffiliation[Also at ]{Center for Semiconductor Physics in Nanostructures}
\affiliation{$^{1}$University of Oklahoma, Department of Physics and Astronomy,
Norman OK 73019}

\date{\today}

\begin{abstract}
We calculate 
the zero temperature electrostatic
properties of charged one and two dimensional arrays of
rings, in the classical and quantum limits.
  Each ring is assumed to be an ideal ring of negligible
width, with exactly one electron on the ring that interacts only with
nearest neighbor rings.  
In the classical
limit we find that if the electron is treated as a point particle,
the 1D array of rings can be mapped to an Ising antiferromagnet,
while the 2D array  groundstate is a four-fold degenerate ``stripe"
phase.  In contrast, if we treat the electrical charge as a
continuous fluid, the distribution will {\em not} spontaneously
break symmetry, but will develop a charge distribution reflecting
the symmetry of the array.  In the quantum limit, the competition
between the kinetic energy and Coulomb energy allows for a
transition between unpolarized and polarized states as a function
of the ring parameters.  This allows for a new class of polarizable
materials whose transitions are based on geometry, rather than a 
structural transition in a unit cell.

\end{abstract}

\maketitle
  Advances in technology have
made it possible to fabricate
structures like dots, and wires on the micro- or nano-scale.
Such quantum dots act as
artificial atoms,
with spectra\cite{QDspectra} and shell structure
similar to those of real atoms.
Recently it has also become possible
to fabricate arrays of rings, each only tens of nanometers across. 
They can be fabricated either
by dry etching\cite{MattRef}
or by using MBE techniques to foster self-assembled
InGaAs/GaAs rings.\cite{SelfAssembly}  
If a quantum dot can be thought of
as an artificial atom, then rings (and other more complicated structures)
can be thought of as ``atoms'' that do not exist in nature. We can vary
their geometry  to control properties such as polarizability, and coupling
strength in order to produce new types of correlated systems.


%

\begin{figure}[t]
\begin{center}\leavevmode
\includegraphics[width=.7\linewidth]{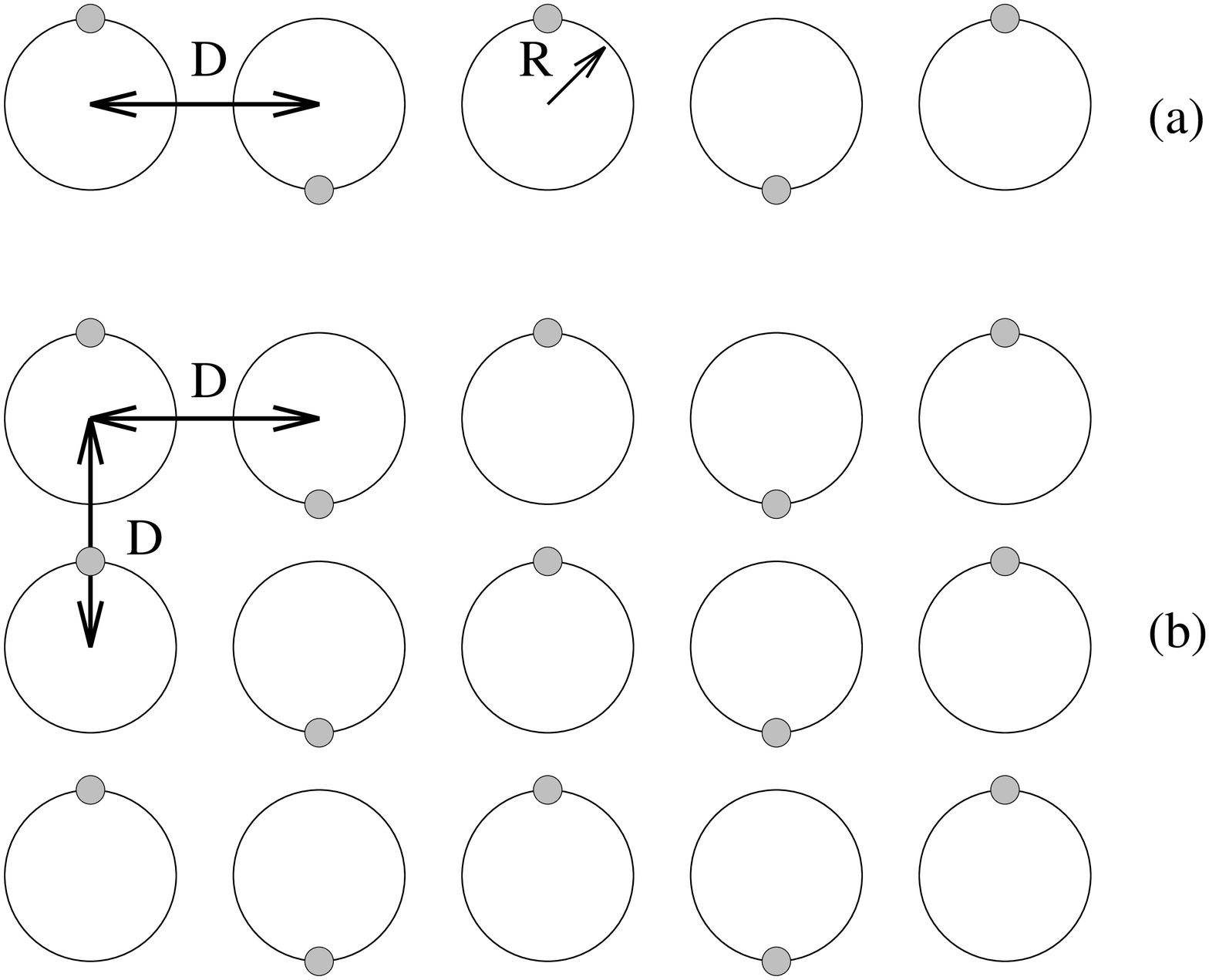}
\noindent
\caption{ 
Figure 1:  A schematic picture of the groundstate of classical point
electrons  for (a) 1D and (b) 2D square
arrays of rings.  The ring radius is $R$ and the separation is $D$.  The
1D ordering is antiferroelectric and thus has a double degenerate
groundstate.  The 2D square array is striped antiferroelectric, and thus
has a fourfold degenerate groundstate. \label{fig1}
}
\end{center}\vspace{-0.5cm}
\end{figure}

In this letter we report the results of analytical and numerical
calculation of polarization of one and two dimensional arrays of singly
charged interacting quantum and classical rings at zero
temperature (fig.(\ref{fig1})). 
 The restriction that each electron resides on a ring 
causes them to exhibit different ground state properties when they are
assembled into arrays of rings
 than are found in arrays of simply connected quantum dots.
We find that the minimum energy configuration of classical point
charges in an array of rings
is an  antiferroelectric  (AFE) pattern in 1D and a striped AFE pattern
in 2D square and triangular
arrays.  If we replace the point charge with an extended classical
charge fluid, we no
longer spontaneously break the symmetry, but instead produce charge
distributions that reflect the symmetry of the array.  In the quantum
limit we find that the groundstate makes a transition from uniform to AFE
order in both 1D and 2D as a function of the array parameters.
In none of our results were globally ferroelectric (FE) order obtained.
We present these findings below, and then conclude with a discussion of
how they might be tested by experiment.


{\em Classical Results:}
In order to gain
intuition for the quantum problem, we will examine the classical
case.  Here there are two limits: discrete charges and continuous charges
We first examine the limit of
ideal point charges on each ring with nearest neighbor
interactions only.  
  
We start by considering a 1D array of ideal rings. The radius of
each ring is $R$, and their center-to-center separation is $D$. On
each ring we place one electron with charge $e$ (fig.(\ref{fig1})).  
Because the electrons are confined to move on a ring, their moment
 is constrained in magnitude but not in direction.

The energy of a 1D array is given by 
$
U_{1D}= \sum_{i=1}^N  {e^2 / \left\vert {\vec r}_i(\theta_i) -
              {\vec r}_{i+1}(\theta_{i+1})\right\vert}$,
where
$\theta_i$ is the location of the $i$-th electron as measured from
the horizontal axis.  In the dipole approximation we can write
this as 
\bea U_{1D}-U_0&\approx& 
{\epsilon^2 \over D} \sum_i ({3\over 2} \cos{2
\theta_i} + \cos{(\theta_i-\theta_{i+1})}  \nonumber \\
&-&3 \cos{(\theta_i+\theta_{i+1})})  \label{u1dtheta}\\ 
&=& 
{\epsilon^2 \over D} \sum_i \,\, \left(2{\vec s}_i
\cdot {\vec s}_{i+1} + {3\over 2} ({\hat D}\cdot
({\vec s}_i-{\vec s}_{i+1}))^2 \right)  \nonumber
\eea 
where $\epsilon\equiv R/D$ and $U_0$ is a constant,
$U_0\equiv{N\over D} \left( 1 + { \epsilon^2 \over 2} \right) $ 
  In the second expression we write the position
of the charge as ${\vec s}_i$, a vector in the 2D plane
pointing from the center of the $i$-th ring to the charge on that
ring.  
The $\cos2\theta$ (or $\vec D \cdot \vec s$)
term explicitly breaks the rotational symmetry, driving
the system from XY to Ising-like behavior.  
The last two terms in eq.(\ref{u1dtheta})  compete: 
the first driving the system to be ferroelectric (like direction) while the
latter, larger term
 favoring states where neighbors point in opposite directions.
Thus the system at zero temperature will order in an
AFE pattern in 1D.  
This is born out by numerical Monte
Carlo simulations of the exact Coulomb interaction, 
which reproduce the AFE pattern on finite
size arrays (Fig.\ref{fig1}).

If we expand the potential to quadratic order in the displacement and
calculate the normal modes we find:
$$
\omega^{(1D)}_\pm (k)= 2 \omega_0\sqrt{4\pm 2\cos{kD\over 2}}
$$
where $\omega_0\equiv  \sqrt{e^2 /m^* D^3}$. 
Both the modes are gapped, since there
is an Ising-like term that provides a harmonic restoring force at each
site.  Note that they are independent of ring radius at this level of
approximation..

We may repeat the analysis above for a 2D array of rings.  
The Coulomb energy for an $N\times N$ array may be written as
\bea
U_{2D}&\approx& {2N^2 \over D} \left(1 + {5\over 4} \epsilon^2 \right)\\
&+&{\epsilon^2\over D}
\sum_{i,j } \left( 2 {\vec s}_{i,j}\cdot{\vec s}_{i+1,j} +
 {\vec s}_{i,j}\cdot{\vec s}_{i,j+1}\right)  \nonumber \\ 
&-&
{3\over 2} \left( s^x_{i,j} s^x_{i+1,j}+ s^y_{i,j} s^y_{i,j+1}
\right)  \nonumber \eea where $i$ and $j$ denote the x and
y indices of the ring in the plane. Since each ring in the plane
(except the ones on the top and right boundaries) has one neighbor
to the right and one above it, this covers the array without
double counting.   If we neglect edge effects the minimum energy
configuration of the system is AFE in one direction (x or y) and
FE in the other (y or x). Since the AFE state is itself doubly
degenerate, we have a four-fold degenerate minimum energy state
(Fig.(\ref{fig1}). 
This  is  supported by numerical simulations which enforce
(unphysical) periodic boundary conditions on the array.  In more
realistic 2D finite arrays the long range order is perturbed by a
``vortex'' like pattern on the boundaries where the charge on each
ring resides on the point that puts it on the outermost  perimeter
of the array.

A normal mode analysis shows that the excitation spectrum is given by
\be
\omega^{(2D)}_\pm (k)= \omega_0\sqrt{6+ 2\cos{k_y D}
\pm 4\cos{k_x D\over 2}}
\ee
where we have assumed that the stripes run vertically in the ordered phase,
as in fig.(\ref{fig1}).  Note that the acoustic mode is now {\it ungapped},
at the wavevector $\vec k_0 = (0,\pi/D) $, in contrast to the 1D case.
This result holds in the case of only nearest neighbor interactions, where
the increase in the energy due to departure from AFE order in the
horizontal neighbors is compensated by the motion of vertical neighbors
from FE allignment.  This mode differs from the magnetoplasmon\cite{magpla}
mode of metal ring arrays.

A second classical limit that contrasts with the point charge model 
is a continuous charge fluid.  That is,
each point on the ring has a charge density $\rho_i(\theta_i)$ such that
the integrated charge density per ring is  a constant,
$\int \rho_i(\theta_i) d\theta_i = 1$.  We seek a minimal solution to the
variational quantity:
\be
I = {1\over 2} \int d\theta \int d\theta^\prime \sum_{i, j}
{\rho_i(\theta) \rho_j(\theta^\prime) \over | {\vec r}_i -{\vec r}_j | } \\
+\lambda  \sum_i \int d\theta \rho_i(\theta) \ee
For a 1D ring this is expression is divergent due to the infinite self
energy.  
We can regularize this in several ways.
One method is to introduce a short distance cutoff $\lambda$ to
the Coulomb interaction, discretize the integral equation, and then
 solve the problem numerically.  This can also be done analytically,
expanding all results to lowest order in the cutoff, $\epsilon$.

\begin{figure}[t]
\begin{center}\leavevmode
\includegraphics[width=.7\linewidth]{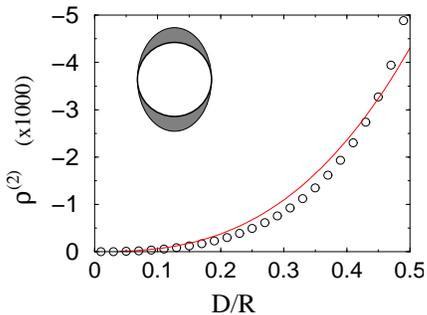}
\noindent
\caption{
A plot of the second fourier amplitude of the classical charge
distribution on a ring in a 1D horizontal
array.  The circles are numerical results, the solid 
line is a scaled  plot of eq.\ref{fourier}.  Scaling is required since the 
analytic result neglects all higher fourier modes.  Inset: a sketch of the
charge distribution that corresponds to this Fourier mode.  Note that the
symmetry of the array is not broken by the charge distribution.
\label{contQ}
}
\end{center}
\vspace{-0.5cm}
\end{figure}

An approximate analytic solution  can then be obtained by Fourier
expanding the distribution,  
keeping
only the first three modes. We find that amplitude non-trivial Fourier mode
as a
function of $\lambda$ and $\epsilon=R/D$ for $\rho$   is given by
\be
\hat\rho^{(2)}\approx \frac{-3\,\pi \,\epsilon^3\,
    \left( 2 - 5\,{\lambda }^2 \right) }{4\,
    \left( -2 + 4\,\log (\epsilon) - 
      4\,\log (\lambda) \right) } \label{fourier}
\ee
We compare this analytic result with the numerical approach in
Fig.(\ref{contQ}).
A second method is to obtain 
analytic results by smearing the ring charge density over a  
very short cylinder, whose height $h$ is small compared to its radius.
Analytic results are in qualitative agreement with the cutoff expansion.

 {\em Quantum mechanical results:} The
quantum mechanical case at first glance seems 
similar to the classical fluid charge
case.  Although there is no  self-interaction, the quantum particle
has a kinetic energy which opposes localization, hence
for low values of interaction strength quantum fluctuations
should destroy the polarization. However for large values of interaction
it is not {\it a priori} obvious 
 whether the polarization pattern beaks  the
symmetry.

\begin{figure}[t]
\begin{center}\leavevmode
\includegraphics[width=.7\linewidth]{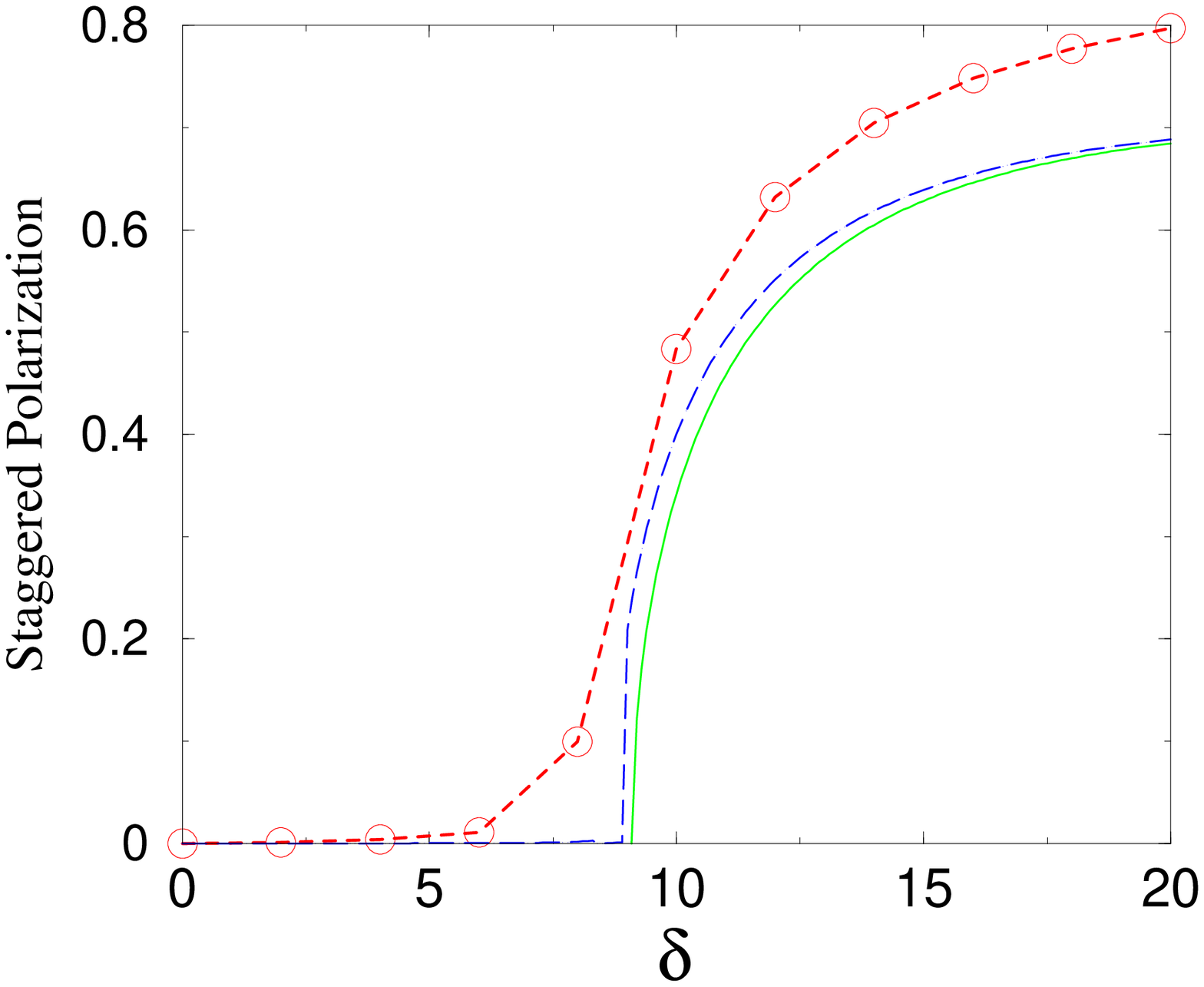}
\noindent
\caption{A comparison numerical and analytical 
calculations
of the staggered polarization as a function of 
$\delta=\left(e^2/ D\right) /\left( \hbar^2/ 2mR^2\right)$ 
in a 1D quantum ring
array obtained in  the Hartree approximation. 
The numerical results are for the case include Fourier modes $|m|\le 1$
(dashed line) and $|m|\le 6$ (circles).  The solid
line is the analytic result assuming $|m|\le 1$.
The quantity $\delta$ is a
measure of the competition between the Coulomb interaction and the quantum
kinetic energy.  \label{hart} }
\end{center}\vspace{-0.5cm}
\end{figure}

Again we assume exactly one charge on each ring and only
nearest neighbor Coulomb interaction. The thickness of each ring is
much smaller than its radius so that the only 
coordinate of the charge is the angular position.  
We decompose the wavefunction in each ring into a limited number of
Fourier modes,
$\psi_i(\theta)=\sum_{n=-n_0}^{n_0}c_n e^{in\theta}$,
 and then
solve the system numerically in the
Hartree approximation.
We use a dimensionless Hamiltonian for the $i$-th ring:
 \be
{\hat h}_i= -{
 }{\partial^2\over\partial\theta_i^2}+\delta\sum_{j}\int_0^{2\pi}{|\psi_j(\theta')|^2\over |\vec{r}_j(\theta
 ')-\vec{r}_i(\theta)|}d\theta'  \label{dimlessH}
 \ee
where the sum is over the nearest neighbors of $i$-th ring and
$\delta=\left(e^2/ D\right) /\left( \hbar^2/ 2mR^2\right)$ is 
the interaction
strength and energy is measured in units of
$ \hbar^2/ 2mR^2$. 
We impose  periodic boundary conditions on the array and by an
iterative self-consistent method we find  that in a 1D array of rings for a
finite value of $\delta=\delta_c$ there is a change of ground
state from totally unpolarized state $|{\vec P}|=0$ to AFE state
$|{\vec P}|=P_0$ where $\vec P$ is the staggered polarization vector ${\vec
P}={1\over N}\sum_i(-1)^i\int {\vec
r}_i(\theta)|\psi_i(\theta)|^2d\theta$ for 1D array. The ground
state energy per ring: 
\bea
 E&=&
{1\over N}\sum_{i}\int\psi_i^\ast(\theta)
{\partial^2\psi_i(\theta)\over\partial\theta_i^2}d\theta \nonumber \\
&+&
{\delta\over 2N}\sum_{\langle i,j\rangle}
\int_0^{2\pi}{|\psi_i(\theta)|^2|\psi_j(\theta')|^2\over |\vec{r}_j(\theta
 ')-\vec{r}_i(\theta)|}d\theta'd\theta \label{gse}
\eea
 also shows a change in behavior at $\delta_c$.  Results of these
calculations are plotted in Fig.(\ref{hart}).

We can obtain analytical insight by
assuming that the ground state wave function for all rings in each
sublattice is the same, so that we can employ a simple variational wave
function for each sublattice $\psi_A(\theta)={\sqrt{1-y^2}\over
\sqrt{2\pi}}+{y\over\sqrt{\pi}}\cos(\theta-\phi)$ and
$\psi_B(\theta)={\sqrt{1-y^2}\over
\sqrt{2\pi}}-{y\over\sqrt{\pi}}\cos(\theta-\phi)$
alternatively. Inserting these functions into (\ref{gse}) and minimizing
the energy using dipole expansion we find an approximate behavior
for energy and polarization.
In the ground state we find  
$\phi={\pi\over 2}$ and 
\bea
 y(\delta,\epsilon)= \cases{
   {1\over 4}\sqrt{11-{4\over\delta\epsilon^2}},& for $\delta \geq
    \delta_c(\epsilon)={4\over 11}\epsilon^{-2} $\cr
    0,& for $\delta\leq \delta_c $ }
\eea 
The numerical Hartree  results for the 
energy and polarization are in agreement with the variational
calculation when  we restrict the number of Fourier modes in the
numerical calculation to $n\in \{-1, 0, 1\}$.
The results are little changed when 
more Fourier modes are introduced to the numerical calculation or 
even when the 
exact Coulomb interaction is used in our numerical code.

Using the
same Hamiltonian as eqn(\ref{dimlessH}) for a
 2D array and with periodic boundary
conditions we are able to find the behavior of the ground state
energy and staggered polarization vector as a function of
interaction strength. Again by using limited number of Fourier
modes for each wave function we find that there is a change
of ground state from totally unpolarized state to a {\em striped}
AFE state in which each column has FE polarization but in opposite
direction of its neighbor column. The staggered polarization
vector for this veritically oriented case is expressed as :${\vec P}=
\sum_{i,j}(-1)^{j}{\vec p}_{ij}/N $ where $j$ counts the columns
and ${\vec p}_{ij}=\int {\vec
r}_{ij}(\theta)|\psi_{ij}(\theta)|^2d\theta$ is polarization
vector of each ring.  The ground state is four-fold degenerate (the states
obtained by flipping every spin and or rotating them all by $\pi/2$ yield
the same energy)  and the U(1) symmetry of unpolarized state has been
broken.

\begin{figure}[t]
\begin{center}\leavevmode
\includegraphics[width=.7\linewidth]{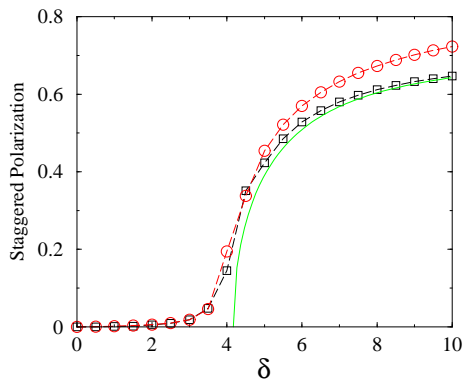}
\noindent
\caption{A comparison numerical and analytical 
calculations
of the staggered polarization as a function of
$\delta=\left(e^2/ D\right) /\left( \hbar^2/ 2mR^2\right)$
in a 2D quantum ring
array obtained in  the Hartree approximation.  
The numerical results are for the case include Fourier modes $|m|\le 1$
(squares) and $|m|\le 6$; the dashed lines are added as a guide.  The solid
line is the analytic result assuming $|m|\le 1$.
The quantity $\delta$ is a
measure of the competition between the Coulomb interaction and the quantum
kinetic energy.  \label{hart2D}}
\end{center}\vspace{-0.5cm}
\end{figure}

Again it is possible to estimate a variational
wave function for each ring assuming the wave functions of the
rings in each sublattice are the same (infinite system or periodic
boundary condition). Using dipole approximation we find
$\phi={\pi\over 2}$ and 
\bea
 y(\delta,\epsilon)= \cases{
   \sqrt{2-{1\over 3\delta\epsilon^2}},& for $\delta \geq
    \delta_c(\epsilon)={1\over 6}\epsilon^{-2} $\cr
    0,& for $\delta\leq \delta_c $}
\eea Numerical Hartree data and variational approximation results
are in good agreement even by using exact Coulomb interaction in
numerical calculations Fig(\ref{hart2D}).

The quantum phase transition in the 1D system should be similar to that of
the transverse field Ising model.\cite{sachdev}  The $\cos{2\theta}$ term
breaks the XY-symmetry so that at low temperatures the system is Ising.
The kinetic energy term tries to delocalize the spin, similar to the transverse
magnetic field in an Ising model.  However, the Coulomb interaction is not
precisely an AFE Ising interaction;  it is a difference of
$\cos{(\theta-\theta')}$ and $\cos{(\theta+\theta')}$ terms that
we can cast it in that form only at
very low temperatures where we assume that $<\theta>$ is sharp on each
ring.  

In 2D there should be a finite temperature phase transition between the
polarized and unpolarized state.  While this Letter focuses only on the
zero temperature behavior, Monte Carlo simulations of the 1+1D quantum
problem not only support the existence of a 1D quantum phase transition,
they document the existence of a 2D transition at finite
temperature.\cite{future}

{\it Considerations for experiment:}  
The parameter $\delta$ determines if
the array will spontaneously polarize;  in 1D the transition is at
$\delta\approx 10$, while in 2D, it is at $\delta\approx 5$.   
It is easy to achieve small values of $\delta$ simply by choosing the ring
separation to be large.  Thus the ``quantum" limit where the kinetic energy
dominates is simple to obtain.  To obtain the antiferroelectrically ordered
state we need large $\delta$.  We may write this as
$\delta=(R^2/a_0 D)\times (m^*/m)$ where $a_0$ is the Bohr radius and $m^*$
is the effective mass of the electron.  
If  $\tilde R$ and $\tilde D$ are $R$ and $D$ measured in 
nanometers, and $\tilde m\equiv (m^*/m)$,
then  $\delta \approx 18.9 (\tilde R^2/\tilde D)\tilde m $.  We require that
the rings do not intersect, so that $\tilde D\geq 2\tilde R$. 
Thus the ability to achieve large values of $\delta$ in 
semiconductors will depend upon the value of the effective mass.  
If we set $\tilde D=2\tilde R$, then 
for GaAs ($\tilde m=0.06$)  1D arrays of 
rings with a radius greater than $\sim$10nm will 
be polarized.  For AlAs ($\tilde m =0.4$)  the crossover radius is about
70nm.  Rings with a smaller radius will not spontaneously polarize, but
instead be isotropic

It is well known that in 1D there is no ordered state for $T>0$
for the Ising model.  However, this is true in the thermodynamic limit.
For small arrays over finite time intervals the system can order.  
If we wish to observe this behavior we want the
characteristic energies of the system to be greater than the temperature.  
For the Coulomb energy $kT<e^2/D$, which we may write as $DT< \rm
1.8\times 10^3$ were $D$ is measured in nm and $T$ in Kelvin.  
For the kinetic energy 
this means $kT<\hbar/2m^*R^2$; if we measure $ (m/m^*) R^2 T>40$ in the same
units.  For GaAs we can choose $R$ to be about 14nm at 4K; choosing
materials with a smaller effective mass or going to lower temperature
allows us to increase the radius.

A AFE polarized ring array will scatter light at a wavelength commensurate
with the inter-ring separation, $D$.  In 1D there is a gap $\sqrt{2}\omega_0$,
which we may write as $2\sqrt{2\tilde m}  (a_0/D)^{3/2}$.   For GaAs rings
with a separation $D=1000$nm this gives $\omega\sim \rm 6.0\times 10^{10}Hz$.
The 2D arrays have a similar sized gap at zone center, but the gap vanishes
at one zone edge.  The excitation spectrum can be probed optically, but
scattering at the edge of the zone is difficult due to the constraints
imposed by conservation of energy and momentum.  Typically in such cases
Raman scattering can be used to investigate the excitations.

Finaly these calculations assume that each ring is singly occupied.  This 
might be
obtained by fabricating the rings upon a thin insulating layer covering a
gate.  By tuning the gate voltage we can bias the system so that it is
energetically favorable for an electron to tunnel to the rings.  The gate
will also serve to cutoff long distance interactions between the rings,
supporting the assumption of the nearest neighbor interactions used here.
Moreover, this letter serves to start investigation into a broad class of
problems, such as rings occupied by an optically excited exciton/hole pair 
or perhaps by a small, varying number of electrons created by a random
distribution of dopants.

The topic of quantum dot arrays and their correlations has obvious and
useful analogies with solid state models of crystalline arrays of atoms.
In this Letter we wish to point out that experimentalists have at their
disposal a host of ``unnatural atoms'' analogs:  rings, quantum dot quantum
wells, quantum rice, etc.  The electrons in these nanoscale
constituents are confined to orbitals that may not have atomic analogs.  
Morever, it may be possible to tune the shape of the constiutent to
optimize some desired collective property such as frustration in electric
or magnetic polarization, high susceptibility or sensitivity to optical
polarization of light.   Even more rich behavior will develop if we allow
electrons to tunnel between these nanoscale periodic structures.

In this Letter we have examined the simple case of arrays of singly charged
classical and quantum rings at zero temperature.  We have found AFE order in 
1D classical systems and a transition to 1D AFE order in the 1D quantum
systems.  In 2D classical and quantum systems there is a similar
striped ordering.  The ordered phases have collective excitations that can
be measured optically.

\begin{acknowledgments}

The authors wish to thank Steve Girvin, Herbert Fertig, and Matthew Johnson
for several useful discussions.
This work is supported by NSF MRSEC DMR-0080054 (MA), and NSF
EPS-9720651 (KM).

\end{acknowledgments}

\end{document}